&pdflatex

\documentclass[10pt,twocolumn,prd,aps,amssymb,amsmath,tightenlines,showpacs]{revtex4}

\usepackage{graphicx}

\newcommand{\cPT}{\ensuremath{\mathcal{PT}}}
\newcommand{\half}{\mbox{$\textstyle{\frac{1}{2}}$}}

\begin{document}

\title{Complex classical motion in potentials with poles and turning points}

\author{Carl M. Bender$^a$}\email{cmb@wustl.edu}
\author{Daniel W. Hook$^{a,b}$}\email{d.hook@imperial.ac.uk}

\affiliation{$^a$Physics Department, Washington University, St. Louis, MO 63130,
USA}
\affiliation{$^b$Theoretical Physics, Imperial College, London SW7 2AZ, UK}

\date{\today}

\begin{abstract}
Complex trajectories for Hamiltonians of the form $H=p^n+V(x)$ are studied. For 
$n=2$ time-reversal symmetry prevents trajectories from crossing. However, for 
$n>2$ trajectories may indeed cross, and as a result, the complex trajectories 
for such Hamiltonians have a rich and elaborate structure. In past work on 
complex classical trajectories it has been observed that turning points act as 
attractors; they pull on complex trajectories and make them veer towards the 
turning point. In this paper it is shown that the poles of $V(x)$ have the
opposite effect --- they deflect and repel trajectories. Moreover, poles shield 
and screen the effect of turning points.
\end{abstract}

\pacs{11.30.Er, 02.30.Fn, 05.45.-a}
\maketitle

\section{INTRODUCTION}
\label{s1}
A $\cPT$-symmetric quantum system is typically governed by a complex
Hamiltonian. Moreover, the boundary conditions on the Schr\"odinger eigenvalue
problem associated with a $\cPT$-symmetric Hamiltonian are imposed in the
complex plane. Thus, when one examines the classical limit of a $\cPT$-symmetric
system, one is led inevitably to the study of complex classical dynamics. A
complex classical dynamical system having one complex degree of freedom is
governed by a Hamiltonian of the form $H(x,p)$, where $H$ is often assumed to be
analytic in both $x$ and $p$ except for isolated singularities or branch cuts in
$x$. For such a system, Hamilton's equations read
\begin{equation}
\dot x=\frac{\partial H}{\partial p},\qquad \dot p=-\frac{\partial H}{\partial
x}.
\label{e1}
\end{equation}

In this paper we use both analytical and numerical techniques to examine the
solutions to these differential equations for complex $x(t)$ and $p(t)$. To do
this we decompose $x(t)$ and $p(t)$ into their real and imaginary parts,
\begin{eqnarray}
x(t)&=&{\rm Re}\,x(t)+i\,{\rm Im}\,x(t),\nonumber\\
p(t)&=&{\rm Re}\,p(t)+i\,{\rm Im}\,p(t),
\label{e2}
\end{eqnarray}
and solve the resulting coupled system of ordinary differential equations.
Complex dynamical systems are interesting in part because the particle
trajectories $x(t)$ typically lie on multisheeted Riemann surfaces. We explore
the qualitative features of the classical particle trajectories in the
complex-$x$ plane, and in particular, we compare the behavior of trajectories in
the vicinity of turning points and in the vicinity of poles of $H$. The
principal result in this paper is that turning points act as attractors
(trajectories are drawn to and pulled around turning points), but poles tend to
repel trajectories and can even screen the effects of turning points.

While complex classical mechanics is a relatively new field there have already
been many papers published on the properties of complex phase space
\cite{R1,R2,R3,R4,R5,R6,R7,R8,R9,R10,R11,R12,R13,R14,R15,R16,R17,R18,R19,R20,R21,R22,R23,R24,R25,R26}.
Most of these studies have focused on Hamiltonians of the form $H(x,p)=p^2+
V(x)$. For such Hamiltonians the complex trajectories associated with a given
energy $E$ cannot intersect. This is because the Hamiltonian is invariant under
classical time reversal $t\to-t$. For a given value of $x$ there are two
possible values of the complex velocity $\dot x$, one corresponding to the
trajectory going forward in time and the other corresponding to the trajectory
going backward in time. However, Hamiltonians such as
\begin{equation}
H(x,p)=p^n+V(x)\quad(n>2)
\label{e3}
\end{equation}
can have trajectories that cross themselves. For a given value of $E$ there may
be $n$ trajectories emanating from a point $x$, each one having a different
complex value of $\dot x$. Furthermore, if $n$ is odd, $H$ is not time-reversal
invariant. This means that if we allow time to run backward, a trajectory will
not retrace itself. Hamiltonians of the form (\ref{e3}) are especially
interesting because, as was shown in Ref.~\cite{R11}, the turning points of
these Hamiltonians deflect nearby trajectories by an angle that depends on $n$;
$n=2$ gives a deflection of $180^\circ$, $n=3$ gives a deflection of
$240^\circ$, $n=4$ gives a deflection of $270^\circ$, and so on.

\begin{figure}
\begin{center}
\includegraphics[scale=0.225]{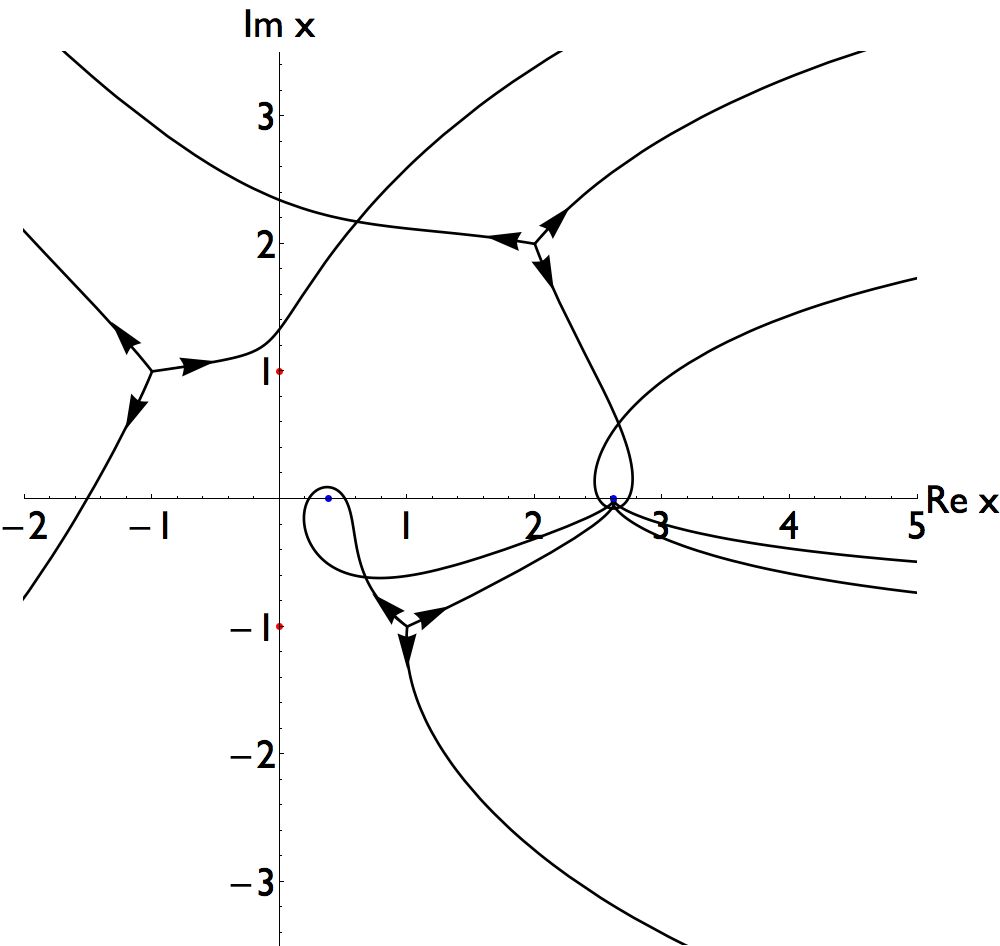}
\end{center}
\caption{Nine complex trajectories associated with the Hamiltonian (\ref{e4}).
All trajectories represent the complex motion of particles having energy $E=
\frac{1}{3}$. There are two turning points at $x=\half\left(3\pm\sqrt{5}\right)$
(blue dots in electronic version) and two poles at $x=\pm i$ (red dots). Because
the Hamiltonian contains a cubic momentum term, at each point in the complex-$x$
plane there are three possible directions for the classical trajectories, and
these directions are separated by $120^\circ$. We have chosen three points at
random, $-1+i$, $1-i$, and $2+2i$, and have plotted the three trajectories
originating from each point. The trajectory that begins at $1-i$ in the
northwest direction encircles both turning points and is deflected by
$240^\circ$ each time. Note that the path does not intersect itself because it 
crosses branch cuts emanating from the turning points and lies on different
sheets of a Riemann surface. The trajectory that begins at $-1+i$ in an eastward
direction is pushed to the north by the simple pole at $x=i$; consequently, the
trajectory is only slightly affected by the turning points.}
\label{F1}
\end{figure}

Figure~\ref{F1} illustrates some of the features of the complex classical 
trajectories discussed in this paper. This figure shows nine trajectories
associated with the Hamiltonian
\begin{equation}
H=p^3+\frac{x}{1+x^2}.
\label{e4}
\end{equation}
For each of these trajectories the classical energy is $E=\frac{1}{3}$. The nine
trajectories emanate from three points in the complex-$x$ plane, $-1+i$, $1-i$,
and $2+2i$. There are two simple poles located at $x=\pm i$ and two turning
points on the real axis located at $x=\half\left(3\pm\sqrt{5}\right)$. The
trajectories enclosing the turning points rotate through an angle of
$240^\circ$. Note that the trajectory emanating from $x=-1+i$ and going eastward
is deflected upward by the pole at $x=i$.

In earlier studies of complex classical mechanical systems many interesting
features were discovered. For example, in Ref.~\cite{R13} it was shown that
a complex classical particle can effectively {\it tunnel} through a potential
barrier on the real axis by following a trajectory in the complex plane that
goes around the barrier. It was also shown that classical mechanics could mimic 
quantum tunneling if the energy had a small imaginary part. This idea was
further developed in Ref.~\cite{R22}. Most recently, Turok showed that these
complex classical trajectories contribute to the saddle point of the functional
integral that represents the tunneling amplitude \cite{R26}.

\begin{figure}
\begin{center}
\includegraphics[scale=0.225]{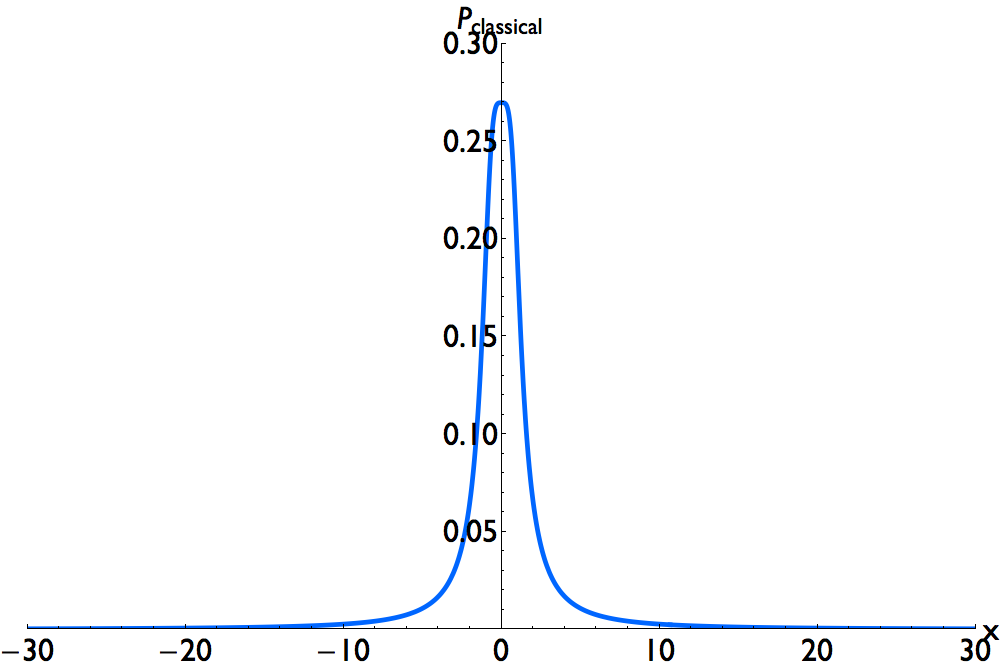}
\end{center}
\caption{Classical probability as shown in (\ref{e7}) for $-30\leq x\leq 30$.
Here, the classical energy is $E=1$.}
\label{F2}
\end{figure}

To illustrate the advantage of complex classical mechanics, we use complex
classical trajectories to explain intuitively how an upside-down
$\cPT$-symmetric potential in quantum mechanics can possess discrete
positive-energy bound states. The Hamiltonian for a quartic oscillator with an
upside-down quartic potential is
\begin{equation}
H=p^2-x^4.
\label{e5}
\end{equation}
The quantum energy levels of this $\cPT$-symmetric Hamiltonian are known
rigorously to be real, positive, and discrete \cite{R27,R28,R29,R30,R31}. The
quantum-mechanical treatment of this system requires that the boundary
conditions on the time-independent Schr\"odinger equation be imposed in Stokes
wedges in the complex plane. To understand the quantum theory heuristically we
examine the classical equations of motion (\ref{e1}) for this Hamiltonian:
\begin{equation}
\dot x=2p,\qquad \dot p=4x^3.
\label{e6}
\end{equation}
For a particle of energy $E$, the classical velocity is $\dot x=2\sqrt{E+x^4}$.
A classical particle is most likely to be found where it is going slowest. Thus,
for real $x$ the {\it normalized classical probability density} $P_{\rm
classical}$, which is the inverse of the particle velocity, is
\begin{equation}
P_{\rm classical}(x)=\frac{2E^{1/4}\sqrt{\pi}}{\Gamma^2(1/4)\sqrt{E+x^4}}.
\label{e7}
\end{equation}
This probability function for $E=1$ is plotted in Fig.~\ref{F2} for real $x$.
Note that the probability density is sharply peaked at $x=0$, which indicates
that the classical particle is most likely to be found near the origin. (It
spends most of its time there.) Thus, we see indications that the quantum
particle may be in a localized bound state.

\begin{figure}
\begin{center}
\includegraphics[scale=0.225]{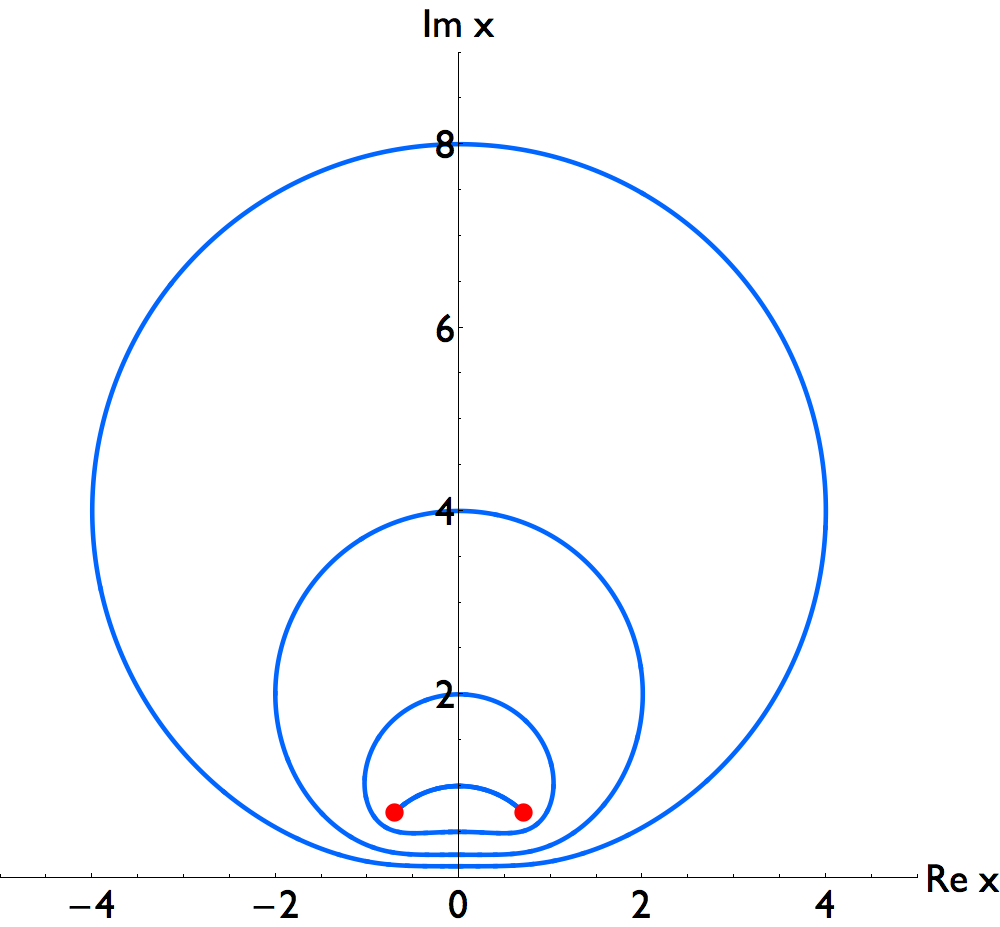}
\end{center}
\caption{Closed periodic trajectories in the upper-half $x$-plane. The energy is
$E=1$. Turning points are located at $x=e^{i\pi/4}$ and $x=e^{3i\pi/4}$ and
are indicated by dots. (The two turning points in the lower-half plane are not
shown.) Initial conditions are $x_0=i,~\frac{i}{2},~\frac{i}{4},~\frac{i}{8}$.}
\label{F3}
\end{figure}

The key to understanding how an upside-down potential confines quantum particles
relies on calculating the time for a classical particle of energy $E$ to travel
to infinity. The time of flight $T$ from $x=0$ to $x=\infty$ is
\begin{equation}
T=\int_{t=0}^T dt=\int_{x=0}^\infty \frac{dx}{\dot x},
\label{e8}
\end{equation}
which is {\it finite}. One may then ask, If the particle reaches infinity in
finite time, where does the particle go next? To answer this question we
investigate the particle motion in the complex-$x$ plane. Four trajectories in
the upper-half $x$-plane are shown in Fig.~\ref{F3}. Note that the trajectories
are all {\it closed} and {\it periodic}. (One of the trajectories is an
oscillatory motion between the turning points at $x=e^{i\pi/4}$ and $x=e^{3i\pi
/4}$.)

\begin{figure}
\begin{center}
\includegraphics[scale=0.225]{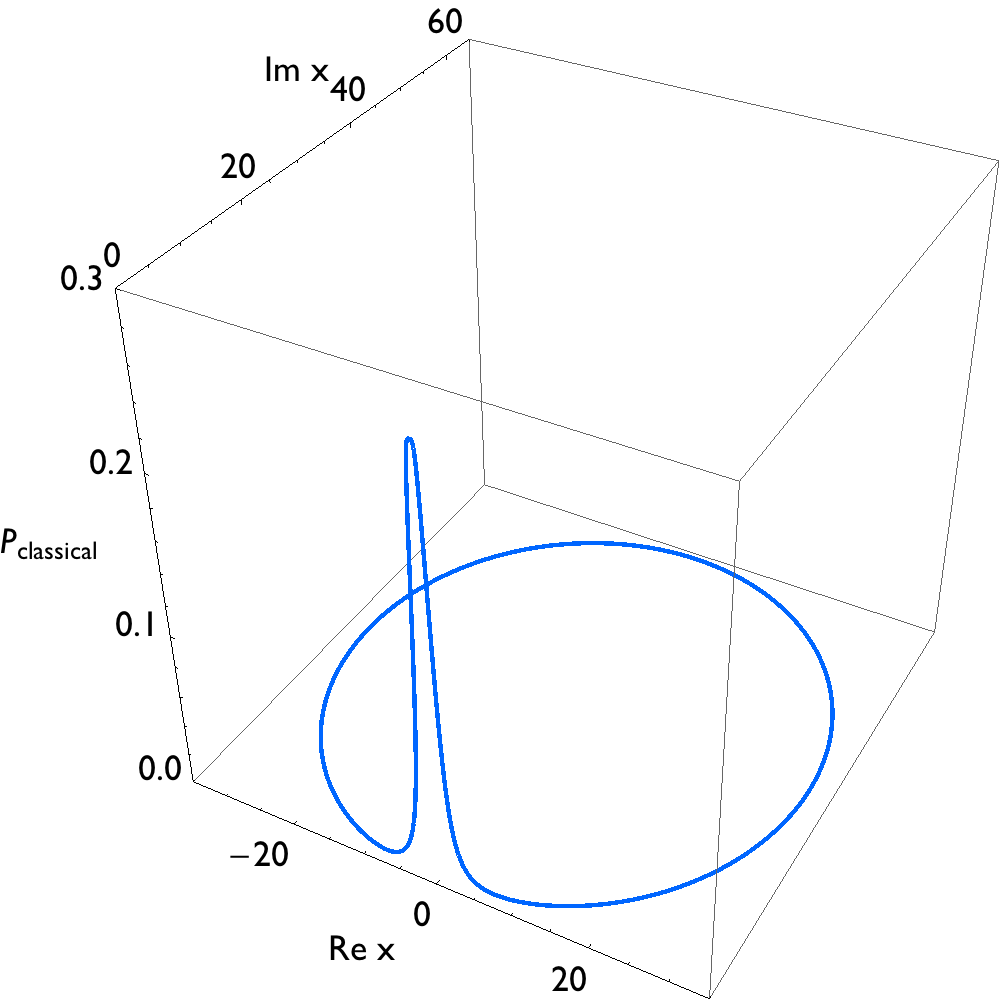}
\end{center}
\caption{Three-dimensional plot of the probability (inverse of the absolute
value of $\dot x$) for a contour beginning at $x_0=\frac{i}{64}$ in the complex
plane. The classical energy is $E=1$. The classical particle executes a rapid
loop in the complex plane but slows down near the origin. Note that the shape of
the peak is the same as that in Fig.~\ref{F2}.}
\label{F4}
\end{figure}

Observe that as the trajectories in Fig.~\ref{F3} approach the real axis, the
classical particle does not simply disappear at $x=\infty$. Rather, as the
classical particle reaches $x=+\infty$, it instantly reappears at $x=-\infty$.
Thus, the classical particle {\it periodically} completes the transit from
$-\infty$ to $+\infty$. (This periodic motion is equivalent to having a source
of particles at $-\infty$ and a sink of particles at $+\infty$.) A
three-dimensional plot of the absolute value of the complex classical
probability is shown in Fig.~\ref{F4}.

To understand why the corresponding quantum energy is discrete rather than
continuous, recall the Bohr-Sommerfeld concept of quantization. We argue that a
quantum particle is a wave, and if the classical motion is periodic, the wave
must interfere with itself constructively. The condition for constructive
interference is
\begin{equation}
\oint dx\,\sqrt{E-V(x)}=\left(n+\half\right)\pi,
\label{e9}
\end{equation}
which is the complex version of the WKB quantization condition. Thus, complex
classical mechanics provides provides the insight to explain how complex quantum
mechanical systems work.

This paper is organized as follows. Section~\ref{s2} explains and illustrates
the nature of complex classical turning points and poles. Then, Sec.~\ref{s3}
examines several examples of Hamiltonians of the form $H=p^2+V(x)$, where the
potential $V(x)$ has poles of order 1, 2, or 3. In Sec.~\ref{s4} we study some
Hamiltonians of the form in (\ref{e3}). Finally, in Sec.~\ref{s5} we make some
brief concluding remarks.

\section{Elementary illustrations of the effects of poles and turning points}
\label{s2}

\begin{figure}
\begin{center}
\includegraphics[scale=0.225, bb=0 0 1000 522]{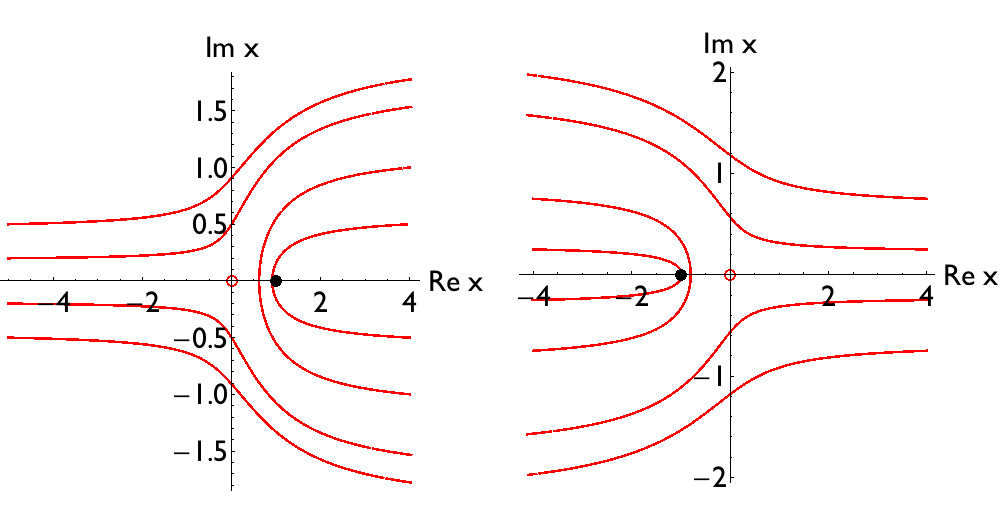}
\end{center}
\caption{Complex classical trajectories for the Hamiltonian (\ref{e10}) with
$n=1$. The potential is $1/x$ (left panel) and $-1/x$ (right panel). The energy
is $E=1$. There is one turning point at $x=1$ (left panel) and $x=-1$ (right
panel). Turning points are designated by filled dots. In both panels the pole at
the origin is indicated by a hollow circle. Left panel: trajectories begin at
$4-\frac{i}{2}$, $4-i$, $-5+\frac{i}{2}$, $-5+\frac{i}{5}$, $-5-\frac{i}{5}$,
$-5-\frac{i}{2}$. Right panel: trajectories begin at $-4+\frac{3i}{4}$,
$-4+\frac{i}{4}$, $4+\frac{3i}{4}$, $4+\frac{i}{4}$, $4-\frac{i}{4}$, $4-\frac{3
i}{4}$. In all cases the turning points attract and deflect trajectories while
the poles repel them.}
\label{F5}
\end{figure}

\begin{figure}
\begin{center}
\includegraphics[scale=0.225, bb=0 0 1000 522]{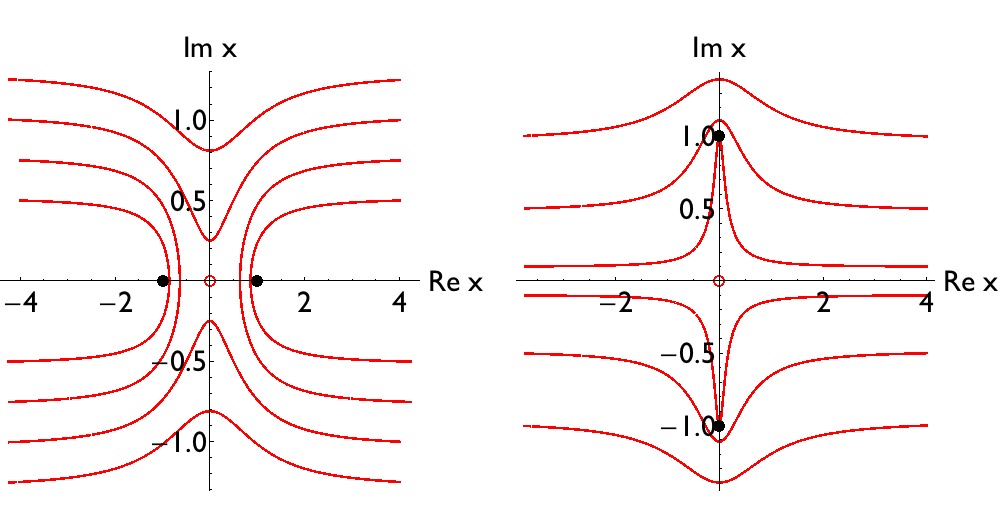}
\end{center}
\caption{Here, the potential is $x^{-2}$ for the left panel and $-x^{-2}$ for
the right panel. For both panels the energy is $E=1$, so there are two turning
points (indicated by filled circles). In both panels the double pole at the
origin is indicated by a hollow circle. Left panel: trajectories begin at $4+
\frac{3i}{4}$, $4+\frac{i}{2}$, $-4+\frac{3i}{4}$, $-4+\frac{i}{2}$, $4+\frac{5i
}{4}$, $5+i$, $4-i$, $4-\frac{5i}{4}$. Right panel: trajectories begin at $4+i$,
$4+\frac{i}{2}$, $4+\frac{i}{10}$, $4-\frac{i}{10}$, $4-\frac{i}{2}$, $4-i$. All
trajectories in the right panel are deflected around one of the two turning
points, while in the left panel the pole overpowers the influence of the turning
points on four of the trajectories.}
\label{F6}
\end{figure}

\begin{figure}
\begin{center}
\includegraphics[scale=0.225, bb=0 0 1000 522]{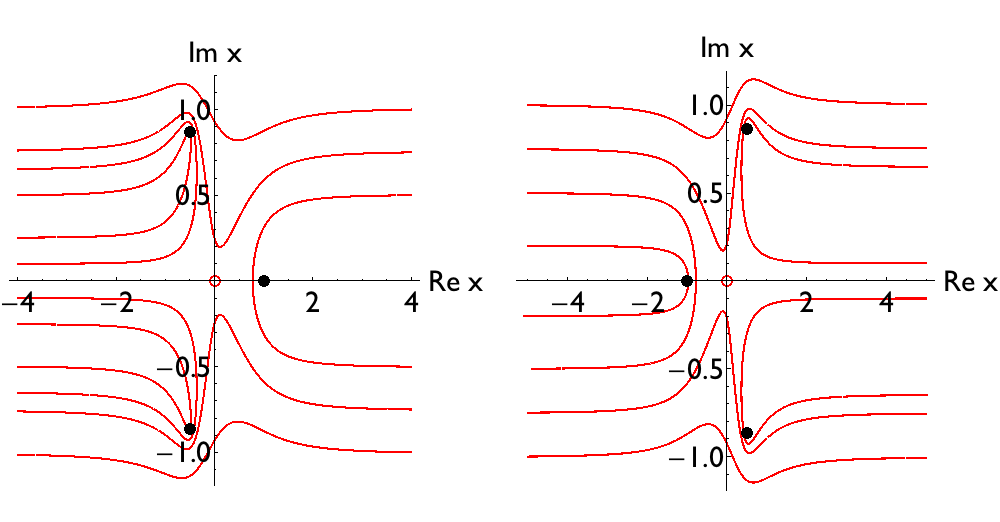}
\end{center}
\caption{Here, the potential is $\pm x^{-3}$ and the energy is $E=1$, so there
are three turning points at $x=1$, $x=e^{\pm2\pi i/3}$ in the left panel and at
$x=-1$ and $x=-1 e^{\pm 2\pi i/3}$ in the right panel. A third-order pole lies
at the origin in both panels. Trajectories in the left panel begin at $-4+\frac{
i}{2}$, $-4+\frac{i}{10}$, $-4-\frac{i}{10}$, $-4-\frac{i}{2}$, $4-\frac{i}{2}$,
$4+i$, $4+\frac{3i}{4}$, $4-\frac{3i}{4}$, $4-i$. Trajectories in the right
panel begin at $-5+\frac{i}{2}$, $-5+\frac{i}{5}$, $5+\frac{i}{10}$, $5-\frac{i}
{10}$, $-5+i$, $-5+\frac{3i}{4}$, $-5-i\frac{3i}{4}$, $-5-i$.}
\label{F7}
\end{figure}

In this section we demonstrate the effects of poles and turning points on
complex classical trajectories by using simple Hamiltonians of the form
\begin{equation}
H=p^2\pm x^{-n}\quad(n=1,\,2,\,3).
\label{e10}
\end{equation}
To our knowledge, complex classical trajectories for such potentials have not
been studied before. (In Ref.~\cite{R17} complex elliptic potentials with double
poles were considered, but the emphasis was on deterministic random walks in a
doubly-periodic complex grid and not on the behavior of trajectories near
poles.) For the Hamiltonians (\ref{e10}) there is one pole of order $n$ at the
origin and $n$ turning points in the complex-$x$ plane. In
Figs.~\ref{F5}--\ref{F7}, we plot complex trajectories for the cases $n=1$, 2,
and 3. The left (right) panel of each figure corresponds to the positive
(negative) sign in $H$. In all cases we can see that turning points attract
trajectories and cause the trajectories to be deflected. However, the poles
repel trajectories and tend to screen out the effects of the turning points. To
understand the dynamics, we note that a classical particle goes slowly in the
vicinity of a turning point, and hence the turning point can have a strong
long-range effect on the trajectory. However, if a pole lies between the
trajectory and the turning point, the turning point has little or no effect.

\section{Potentials with two simple poles}
\label{s3}

We have examined in greater detail two potentials, $V_1(x)$ and $V_2(x)$, each
having two simple poles:
\begin{equation}
V_1(x)=\frac{x}{x^2+1},\qquad V_2(x)=\frac{ix}{x^2+1}.
\label{e11}
\end{equation}
In the two subsections that follow we describe the behavior of trajectories
in proximity to separatrices and we study Zeno-type behavior (trajectories
that approach turning points as $t\to\infty$). 

\subsection{Separatrix and Zeno behavior for $V_1(x)$}
\label{ss3A}

\begin{figure}
\begin{center}
\includegraphics[scale=0.225, bb=0 0 1000 948]{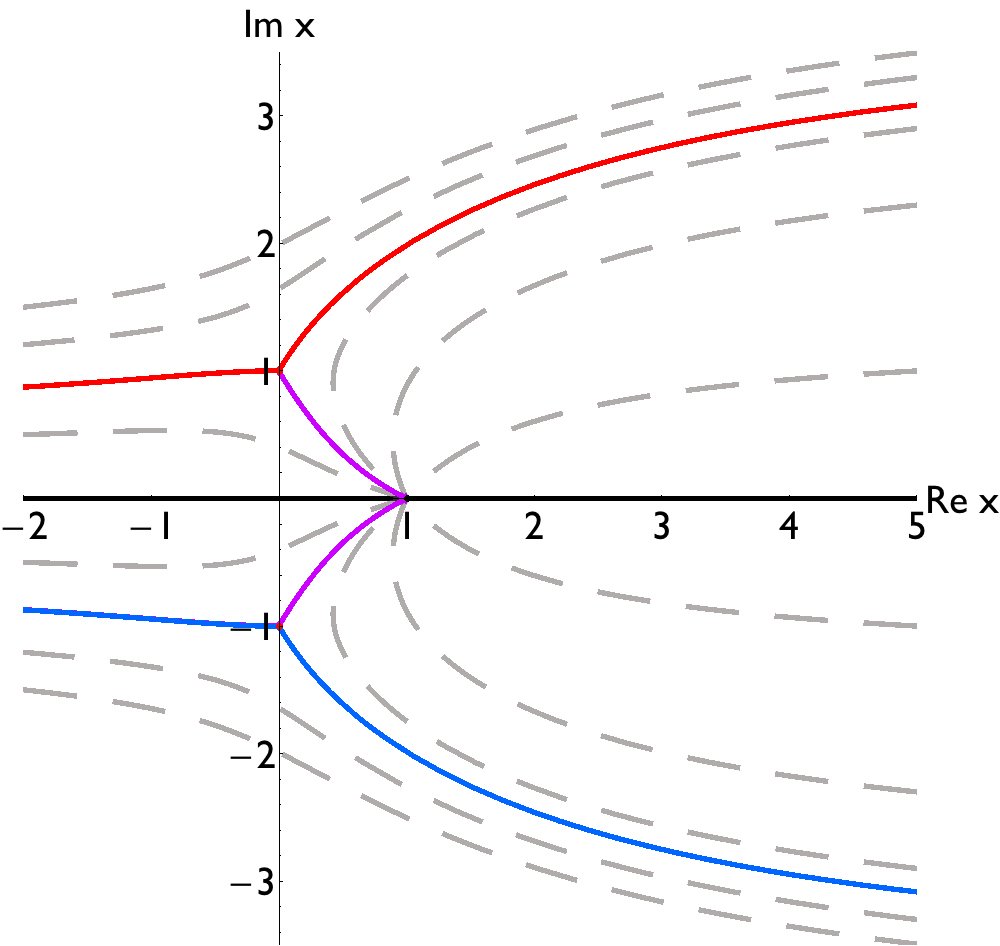}
\end{center}
\caption{Complex classical trajectories for the potential $V_1(x)$ in
(\ref{e11}). The energy is $E=\frac{1}{2}$, so there is just one turning point
at $x=1$ and two poles at $x=\pm i$. Twelve trajectories (dashed lines) and
separatrices (solid lines) are shown. A separatrix emerges from $x=i$ at
$60^\circ$, $-60^\circ$, and $180^\circ$. Trajectories start at ${\rm Re}\,x=-
2$, and ${\rm Im}\,x=0$, $\pm\frac{1}{2}$, and $\pm\frac{3}{2}$. Trajectories
also start at ${\rm Re}\,x=5$ and ${\rm Im}\,x=\pm3.3$, $\pm2.9$, $\pm2.3$, $\pm
1$, and $0$. Three branches of the separatrix curve are shown in both the
upper-half and the lower-half plane. The separatrices intersect at $120^\circ$
angles at the poles. The separatrices end at the Zeno point (the turning point)
$x=1$, and they leave the plot at $x=-2\pm0.87i$ and at $x=5\pm3.08i$.}
\label{F8}
\end{figure}

If we choose the energy to be $E=\half$ for the Hamiltonian $H=p^2+V_1(x)$,
there is just one turning point at $x=1$, which is a {\it double} turning point;
that is, a coalescence of two simple turning points as $E$ approaches $\half$.
There are two poles at $\pm i$. Figure~\ref{F8} illustrates a bifurcation and a
complex Zeno effect. We observe the following: The solid curves in the figure
are separatrices. If a trajectory begins between the upper and lower solid
curves (red and blue in the electronic version), it approaches the double
turning point at $x=1$ on the real axis as $t\to\infty$. However, trajectories
that begin above the upper separatrix or below the lower separatrix are repelled
by the poles at $\pm i$, move off to ${\rm Re}\,x=\pm\infty$ as $t\to\infty$,
and are never captured by the turning point. The two separatrix curves that
connect the poles to the turning point distinguish between trajectories that
begin at ${\rm Re}\,x=+\infty$ and at ${\rm Re}\,x=-\infty$.

We can perform the following asymptotic analysis valid near the pole at $x=i$:
Let $x(t)=1+\epsilon(t)$. Then the equation of motion becomes approximately
$$[\epsilon'(t)]^2\sim -2/\epsilon(t).$$
When we integrate this equation and assume that $\epsilon=0$ at $t=0$, we get
$$\epsilon(t)\sim 3^{2/3}2^{-1/3}t^{2/3}e^{i\theta},$$
where $\theta=\pm\frac{\pi}{3},\,\pi$. Thus, near the pole the separatrix splits
into three lines separated by $120^\circ$. (This kind of behavior is reminiscent
of the Stokes structure of Airy functions.) These features are shown in
Fig.~\ref{F8}. 

To perform an asymptotic study for large $t$ we argue as follows. Since $x'(t)=
\partial H/\partial p=2p$, we get 
$$\frac{1}{4}[x'(t)]^2+\frac{x}{1+x^2}=\frac{1}{2}.$$
Therefore,
$$\int dx\,\frac{\sqrt{1+x^2}}{x-1}=t\sqrt{2}+C,$$
where $C$ is a constant to be determined by the initial conditions. An exact
evaluation of the integral gives
\begin{eqnarray}
&&\!\!\!\!\!\!\!\!\!\!\!\!\!\!\!\sqrt{1+x^2}+\sqrt{2}\log\frac{x-1}{x+1+
\sqrt{2+2x^2}}\nonumber\\
&&\quad +\log\left(x+\sqrt{1+x^2}\right)=t\sqrt{2}+C.
\label{e12}
\end{eqnarray}

To leading order $x\sim t\sqrt{2}$ as $t\to\infty$. However, a higher-order
asymptotic approximation is
$$x\sim t\sqrt{2}-\log t+\sqrt{2}\log(1+\sqrt{2})-\frac{3}{2}\log 2+K,$$
where
\begin{eqnarray}
K&=&\sqrt{1+x_0^2}+\sqrt{2}\log\frac{x_0-1}{x_0+1+\sqrt{2+2x_0^2}}\nonumber\\
&&+\log\left(x_0+\sqrt{1+x_0^2}\right).
\label{e13}
\end{eqnarray}
We have performed a numerical calculation of the complex value of $x$ on the
separatrix at $t=500$ and we have compared this result with the asymptotic
analysis above. At $t=500$ and for $x_0=1.1i$ we get
$$x_{\rm asymptotic}=701.113+3.8073i,$$
while
$$x_{\rm numerical}=701.124+3.8018i.$$
Evidently, the asymptotic analysis is highly accurate.

\subsection{Phase transition behavior for $V_2(x)$}
\label{ss3B}

\begin{figure}
\begin{center}
\includegraphics[scale=0.225, bb=0 0 1000 995]{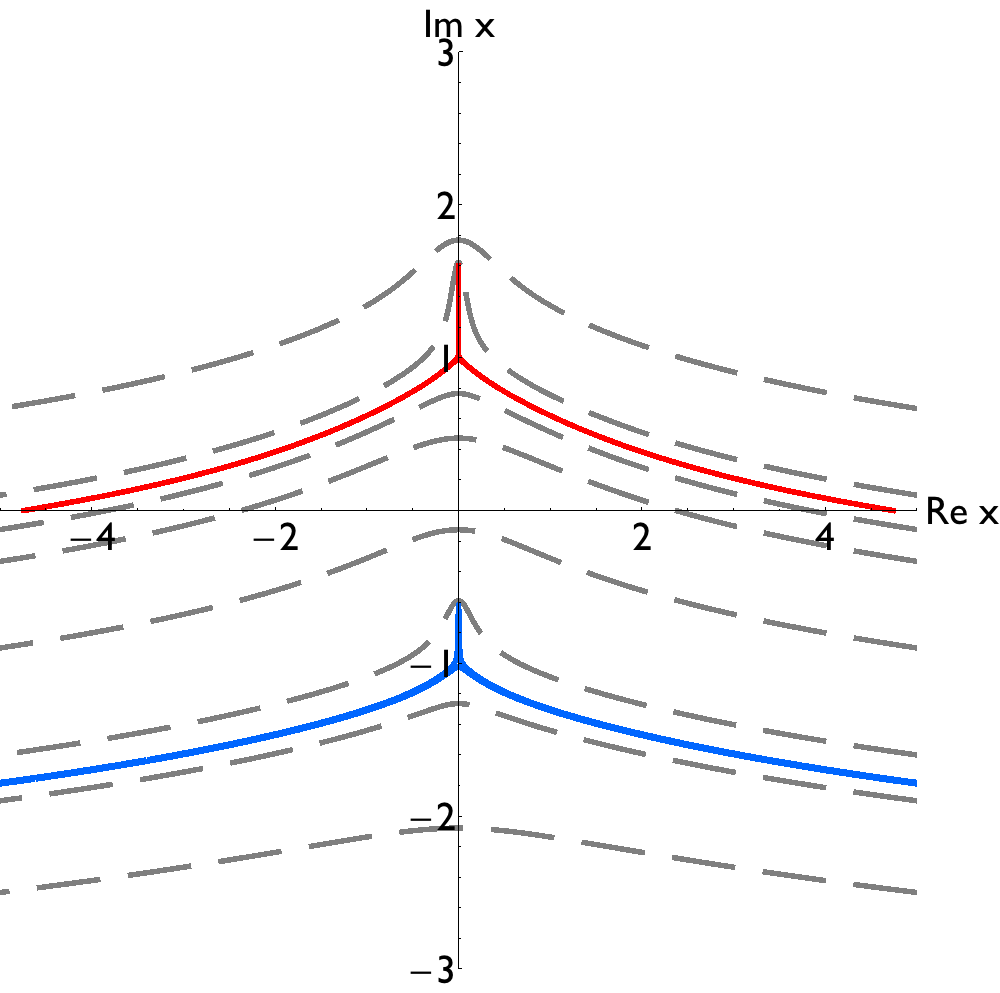}
\end{center}
\caption{Complex trajectories (dashed lines) of a classical particle of energy
$E=1$ in the potential $V_2(x)$. Four paths start at $5+\frac{2i}{3}$ and $5+
\frac{i}{10}$ just above, and at $5-\frac{i}{8}$ and $5-\frac{i}{3}$ just below
the upper separatrix (solid line), which starts on the real axis at $4.735\pm
0.005$. Four more paths begin at $5-\frac{9i}{10}$, $5-\frac{8}{5i}$ just above,
and $5-\frac{19i}{10}$ and $5-\frac{5i}{2}$ just below the lower separatrix
(solid line), which starts at $5-1.785\pm0.005i$. The poles are at $\pm i$ and
the turning points are at $\half\big(1\pm\sqrt{5}\big)i$. Trajectories below the
lower separatrix are hardly affected by the turning points. Trajectories between
the separatrices feel the effect of the lower but not the upper turning point.
Trajectories above the upper separatrix only feel the effect of the upper
turning point. Thus, trajectories just above a separatrix have a longer travel
time than those just below a separatrix.}
\label{F9}
\end{figure}

The potential $V_2(x)$ in (\ref{e11}) is $\cPT$-symmetric. Consequently, the
complex trajectories are left-right symmetric \cite{R1}. They start on the real
axis or in the complex plane to the right of the imaginary axis and extend to
mirror-image points to the left of the imaginary axis. We have studied the
complex dynamical system described by the Hamiltonian $H=p^2+V_2(x)$ and we have
chosen the energy of the particle to be $E=1$. Two turning points are located on
the imaginary axis at $\half\big(1\pm\sqrt{5}\big)i$ and two poles are located
on the imaginary axis ar $\pm i$. In summary, on the imaginary axis starting
from below, there is a pole at $-i$, a turning point at $-0.618i$, a pole at
$i$, and a turning point at $1.618i$. (See Fig.~\ref{F9}.)

\begin{figure}
\begin{center}
\includegraphics[scale=0.225, bb=0 0 1000 997]{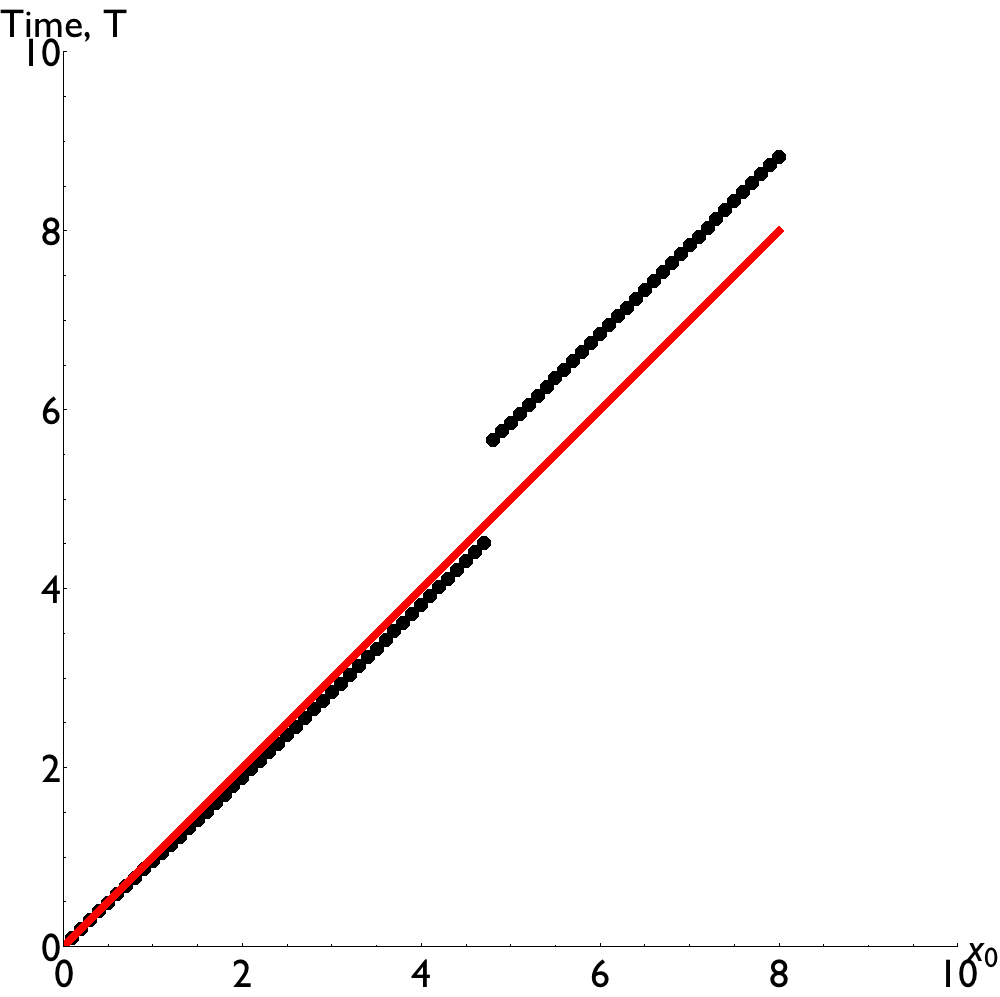}
\end{center}
\caption{Transit time for a particle of energy $E=1$ to go from an initial point
$x_0$ on the positive-$x$ axis to its mirror-image point $-x_0$ on the
negative-$x$ axis. The solid line (red in electronic version) shows the transit
time for a {\it free} particle of energy $E=1$ to travel in the potential $V(x)=
0$. The dotted line (black in electronic version) is the transit time in the
potential $V_2(x)$. The discontinuity is due to the particle going either below
or above the pole at $i$. In the latter case the particle is pulled up and
around the turning point at $1.618i$. In the former case the pole screens the
trajectory of the particle from the effect of the turning point.}
\label{F10}
\end{figure}

We have calculated the transit time numerically from a point in the right-half
plane to its mirror image in the left-half plane. There is a discontinuous
(first-order) transition in this transit time. Below the separatrix curves the
time is shorter than above. This discrete jump is due to the particle going just
below or just above the poles at $\pm i$. The time differences of paths near the
upper separatrix are illustrated in Fig.~\ref{F10}.

To understand the discontinuous curve in Fig.~\ref{F10}, note that below the
pole at $i$ the particle smoothly travels from an initial point on the positive
real axis to its mirror point on the negative real axis. (The trajectory does
not stop there.) The trajectory curves downward in response to the lower turning
point. The high point of the trajectory passes just below the pole at $i$ (and
is horizontal there) but the trajectory is unaffected by the turning point.
Rather, its gentle curvature is due to the more distant lower turning point at
$-0.618i$. The pole at $i$ screens the turning effect of the upper turning
point. However, above the separatrix the trajectory passes just above and is
strongly attracted by the upper turning point. So, the trajectory veers steeply
upward, makes a sharp U-turn around the turning point, and goes downwards in a
mirror-symmetric fashion.

At the separatrix bifurcation the trajectory goes exactly up the imaginary axis
from $i$ to $\half\big(1+\sqrt{5}\big)i=i\phi=1.618i$ and back down to $i$.
Using Hamilton's equation $\dot x(t)=\frac{\partial H}{\partial p}=2p$, we
find that the time $T$ to do this is
\begin{eqnarray}
T &=& 2\int_{x=i}^{i\phi}dt=2\int_{x=i}^{i\phi}\frac{dx}{\dot x(t)}\nonumber\\
&=& \int_{x=i}^{i\phi}\frac{dx}{\sqrt{1-ix/(x^2+1)}}\nonumber\\
&=& 2\int_{s=1}^{\phi}ds\,\sqrt{\frac{s^2-1}{1+s-s^2}}\nonumber\\
&=& 1.05659994.
\label{e14}
\end{eqnarray}

\begin{figure}
\begin{center}
\includegraphics[scale=0.225, bb=0 0 1000 920]{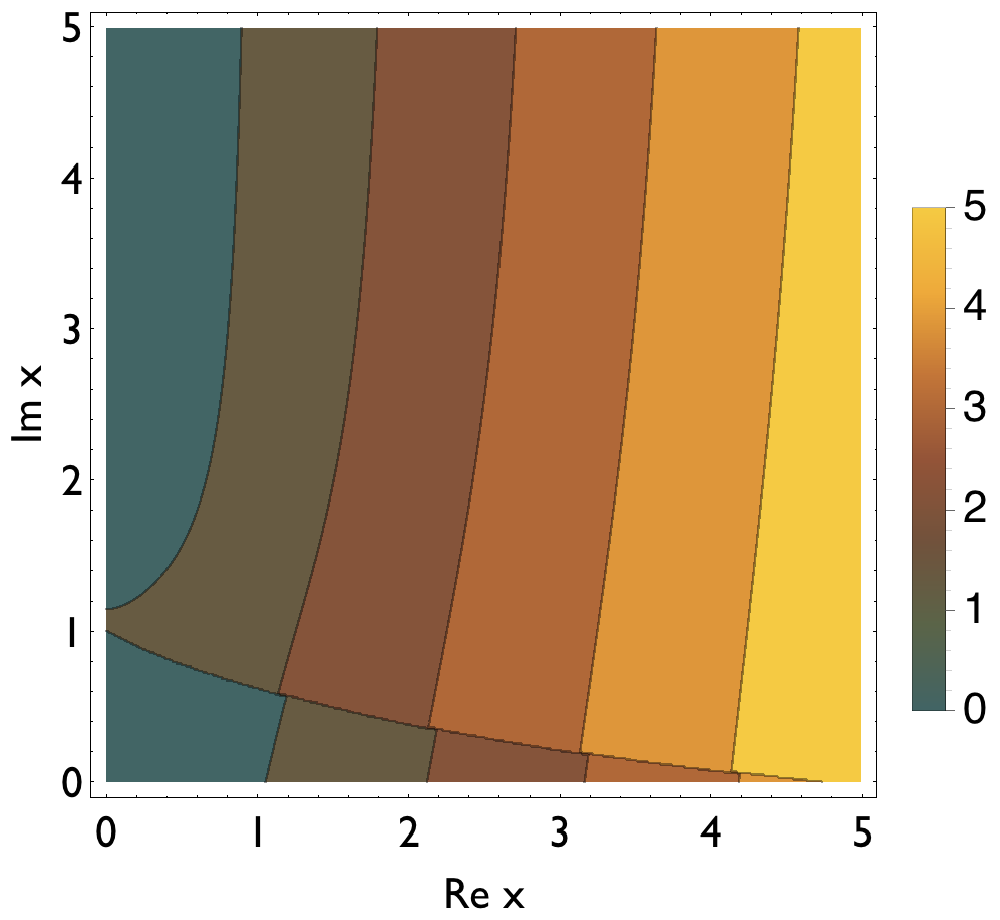}
\end{center}
\caption{Contour plot generalizing Fig.~\ref{F10} from the real axis into
the complex plane.}
\label{F11}
\end{figure}

While Fig.~\ref{F10} plots the transit time for a trajectory beginning and
ending on the real axis, Fig.~\ref{F11} extends these numerical results into
the complex plane. In the positive quadrant each complex pixel $x$ satisfying
$0\leq{\rm Re}\,x\leq5$ and $0\leq{\rm Im}\,x\leq5$ is studied. The transit time
is calculated and a color is assigned to indicate the transit time. The phase
transition appears as the boundary curve sloping downward to the right.

\section{Trajectories for $H=p^3+V_1(x)$}
\label{s4}

Complex trajectories for Hamiltonians having higher powers of $p$ have a rich
and elaborate structure. Some previous studies were done in Ref.~\cite{R11} but
these studies focused on pairs of isospectral trajectories having identical
periods. The current paper presents two results concerning Zeno behavior and
bifurcation for the system described by the Hamiltonian $H=p^3+V_1(x)$.

Recall from Sec.~\ref{s1} that the phase space for such a $p^3$ Hamiltonian is
very different from that of a $p^2$ Hamiltonian. The latter case is symmetric
under classical time reversal $t\to-t$ and any point on a classical particle
trajectory in the complex-$x$ plane is associated with two values of $\dot x$
pointing in opposing directions. One value corresponds to the forward time
direction and the other to the backward time direction. However, in a $p^3$
model each point on a complex trajectory is associated with three velocities
oriented at $120^\circ$ to each other, and the correspondence with forward time
evolution and backward time evolution is lost.

\begin{figure}[h!]
\begin{center}
\includegraphics[scale=0.225]{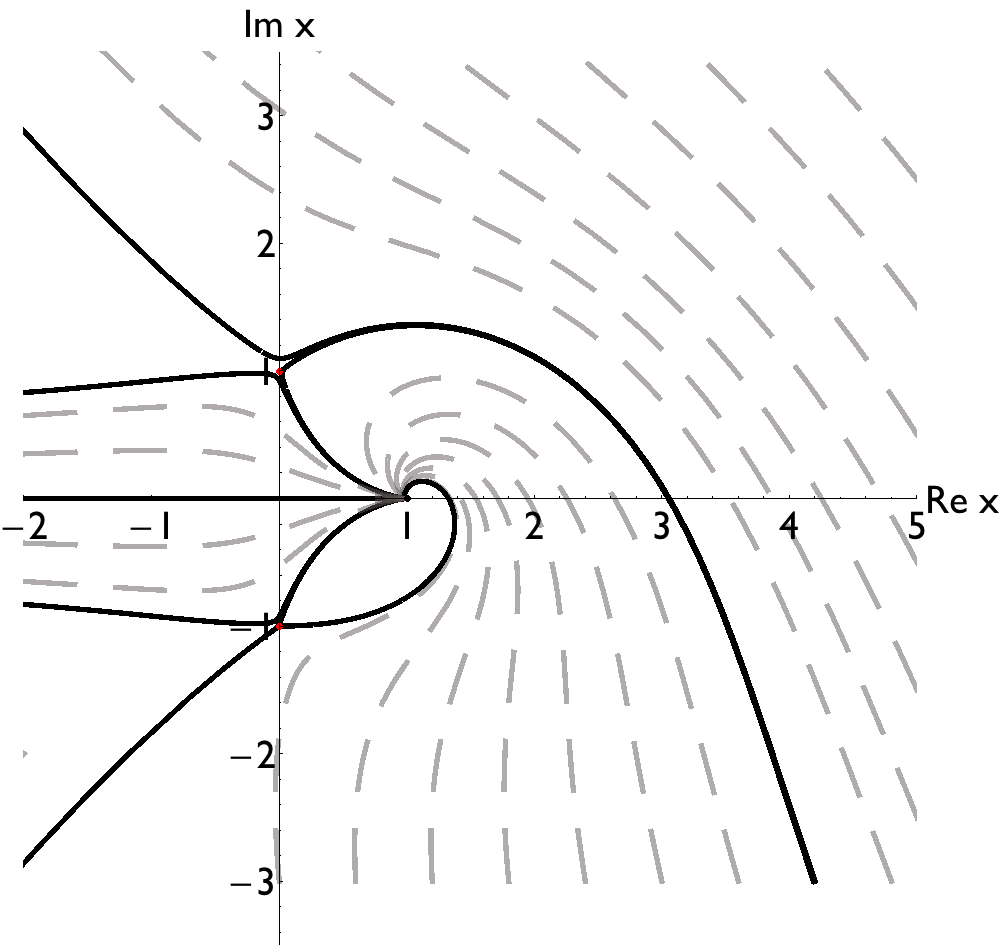}
\end{center}
\caption{Classical trajectories (dashed lines) for the Hamiltonian $H=p^3+V_1(x
)$ for particles of energy $E=\half$. The behavior of trajectories near the pole
at $x=i$ is emphasized. There are three branches of the separatrix curve (solid
lines), which are separated by $120^\circ$.}
\label{F12}
\end{figure}

Figure~\ref{F12} displays 18 trajectories (dashed lines) for a complex classical
particle of energy $E=\half$ governed by the Hamiltonian $H=p^3+V_1(x)$. There
are two poles on the imaginary-$x$ axis at $x=\pm i$. A double turning point is 
situated at $x=1$. Note that the structure of the trajectories is reminiscent of
those shown in Fig.~\ref{F8}. The separatrix paths (solid lines) are now bent
around to create what appears to be an enclosed area in the lower-half plane
between the turning point and the lower pole. The turning point rotates
classical particle trajectories in a manner that is more extreme than in a
$p^2$ theory.

\begin{figure}
\begin{center}
\includegraphics[scale=0.225]{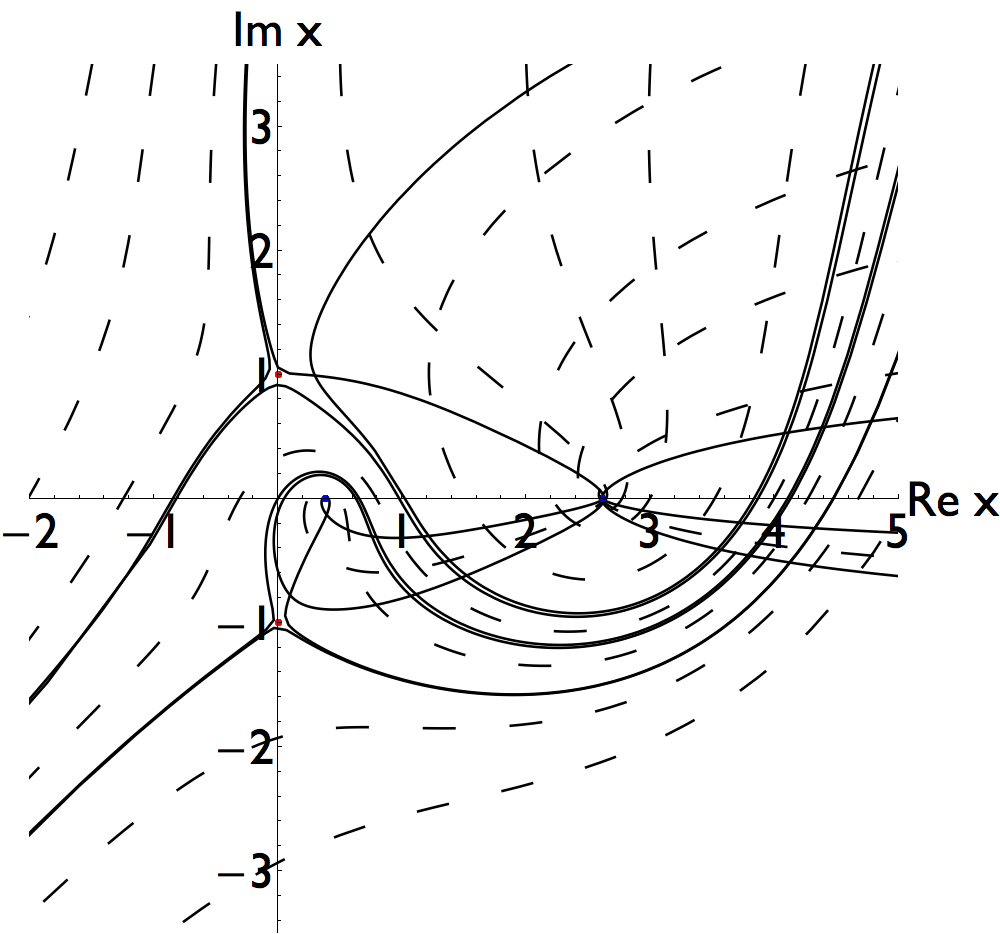}
\end{center}
\caption{Trajectories of a complex classical particle acting under the influence
of the Hamiltonian $H=p^3+V_1(x)$ (dashed lines). The energy of the particle is
$E=\frac{1}{3}$. There are two poles at $x=\pm i$ and two turning points at $x=
\half\big(3\pm\sqrt{5}\big)$. This figure is analogous to Fig.~\ref{F12} except
that we have purposely chosen different directions for the velocities at the
initial points. In the numerical analysis we have required that a path be smooth
and not have elbows. Solid lines are separatrix paths that isolate different 
behaviors. Five different behaviors seen in this complicated figure are depicted
separately in the five panels of Fig.~\ref{F14}.}
\label{F13}
\end{figure}

\begin{figure}[h!]
\begin{center}
\includegraphics[scale=0.5]{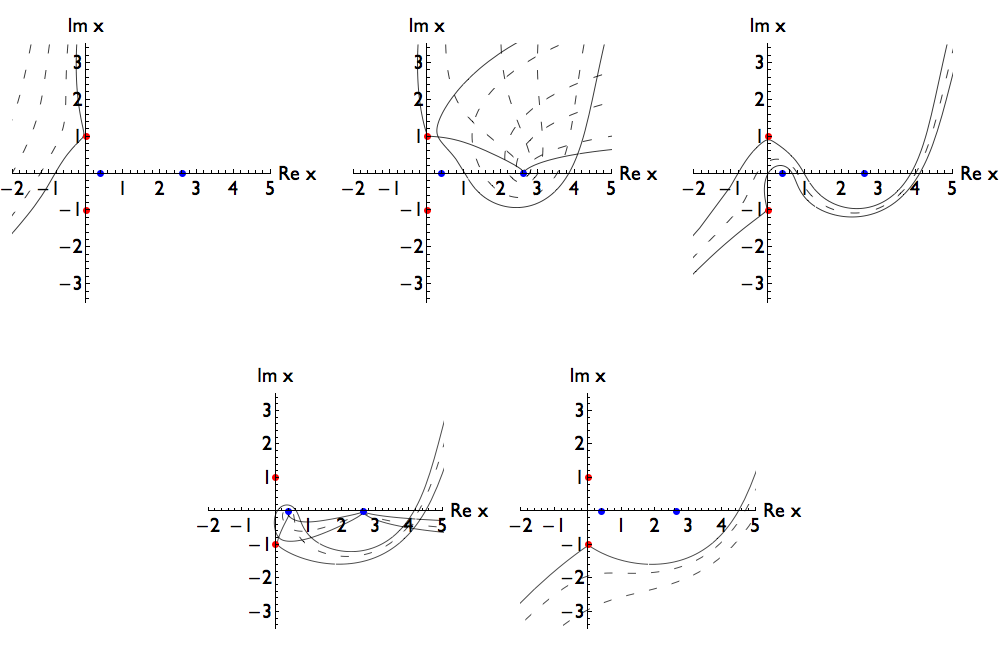}
\end{center}
\caption{Same as Fig.~\ref{F13} but trajectories (dashed lines) are decomposed
into regions between separatrices (solid lines). First panel on top row: All
trajectories originate in the north and travel downward. The poles completely
screen the trajectories from the turning points on the real-$x$ axis. Middle
panel on top row: All trajectories originate in the north and travel downward
while under the control of the right turning point. Each trajectory turns
through an angle of $240^\circ$ onto a different sheet of the Riemann surface
from that on which it entered the plot, and leaves the plot toward the east.
Right panel on top row: A single trajectory shown between the two separatrix
paths. The trajectory is turned by the right turning point but does not encircle
it. Instead it comes under the control of the left turning point but does not
encircle it either as it is repelled by the pole on the negative imaginary axis.
Lower left panel: A trajectory starts in the northeast of the plot and goes in a
downward direction. It is turned by the right turning point but is not close
enough to encircle the turning point. It is then deflected by the pole at $x=-i$
and comes under the control of the left turning point, which it encircles. The
trajectory then emerges from the left turning point on a different sheet of the
Riemann surface and is instantly controlled by the right turning point, which it
also encircles. The trajectory emerges from the right turning point on yet
another Riemann sheet and exits the plot in an eastward direction. Bottom right
panel: Two trajectories start on the right, are partially controlled by both
turning points but are deflected by the pole before being captured, and instead
exit the plot to the southwest.}
\label{F14}
\end{figure}

\begin{figure}
\begin{center}
\includegraphics[scale=0.225]{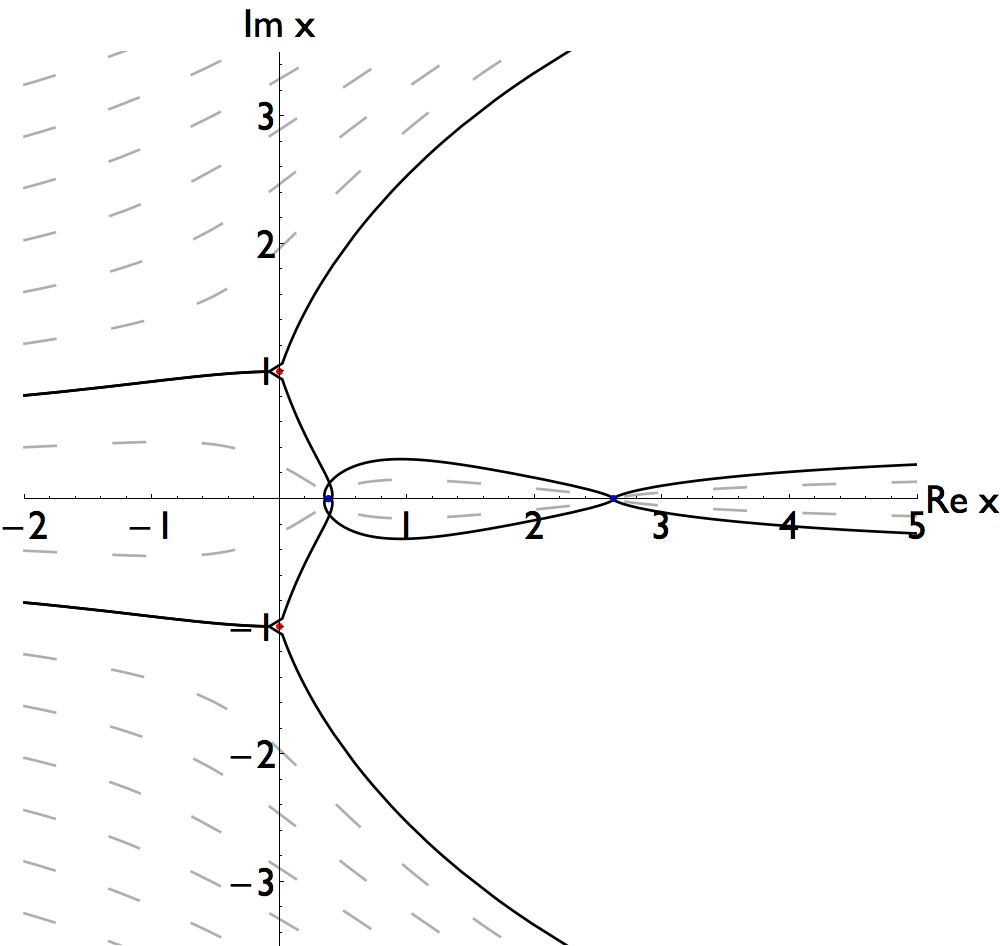}
\end{center}
\caption{As in Figs~\ref{F13} and \ref{F14} but for a different set of initial
conditions for the separatrix curves (solid lines). The trajectories have been
chosen to be consistent with the initial directions of the separatrix curves.}
\label{F15}
\end{figure}

Figure~\ref{F13} displays the trajectories for the same Hamiltonian that was
used to generate Fig.~\ref{F12}, but now $E=\frac{1}{3}$. The trajectories in
Fig.~\ref{F13} are topologically complicated, so we have separated them into
five components as shown in the five panels of Fig.~\ref{F14}. Figure~\ref{F15}
shows trajectories arising from different choices of initial velocities than
those used in Fig.~\ref{F13}.

% \begin{widetext}

% \end{widetext}

\section{Concluding remarks}
\label{s5}

We have examined Hamiltonians whose potentials have poles and zeros. We have
seen that turning points attract trajectories while poles repel trajectories.
We have also seen that trajectories in the complex-$x$ plane can be extremely
complicated for Hamiltonians in which the power of $p$ is greater than $2$, as
shown in Figs.~\ref{F13}--\ref{F15}.

\begin{figure}
\begin{center}
\includegraphics[scale=0.225]{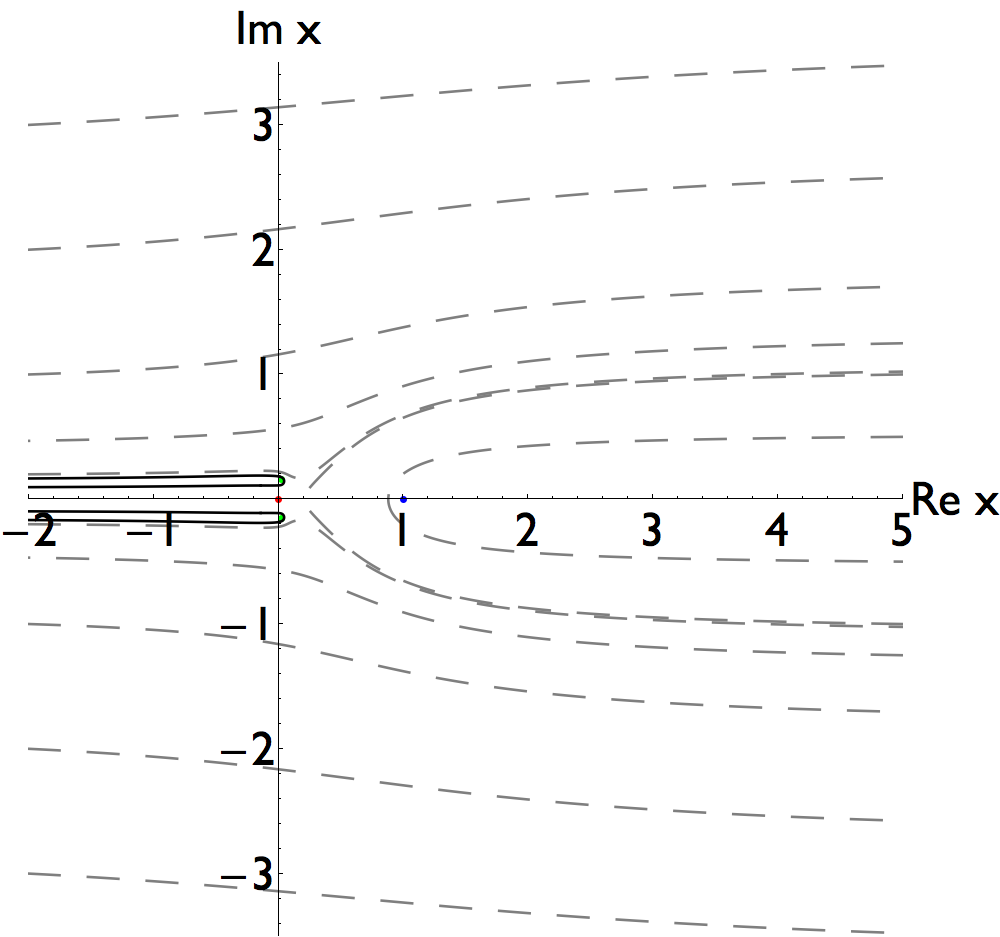}
\end{center}
\caption{Complex classical trajectories for the Hamiltonian $H=p^2+V(x)$, where
$V(x)=e^{1/x}$ and $E=e$. There is a turning point at $x=1$. Note that the
essential singularity at $x=0$ is {\it both} attractive and repulsive depending
on whether the particle approaches the origin from right-half or the left-half
plane. There is one turning point at $x=1$ and an infinite sequence of complex
turning points $x_n$ approaching the origin at ${\rm arg}\,x_n\to\pm\frac{\pi}
{2}$ as $|n|\to\infty$.}
\label{F16}
\end{figure}

Things become even more interesting if the potential has an essential
singularity, and such potentials should be examined in future studies. For
example, consider the potential $V(x)=e^{1/x}$, which has an essential
singularity at the origin, and take the Hamiltonian $H=p^2+V(x)$ and energy $E=
e$. There is a turning point at $x=1$ but there is also an infinite sequence of
turning points at $x_n=r_ne^{i\theta_n}$, where $r_n$ and $\theta_n$ satisfy the
equations $\theta_n={\rm arctan}(2n\pi)$ and $r_n=\cos(\theta_n)$. Thus, as $n
\to\pm\infty$ there is an infinite sequence of turning points symmetrically
placed above and below the origin with a limit point at the origin near $\pm
\frac{\pi}{2}$. For $n=1$ the turning point $x_1$ lies at $r_1=0.157177$ and
$\theta_1=1.412965$ (an angle of $80.9569^\circ$). Figure \ref{F16} displays
some complex classical trajectories for this potential. This figure shows that
the essential singularity is attractive for trajectories approaching the
essential singularity from the right and is repulsive for trajectories
approaching from the left. Trajectories making U-turns around the turning point
at $x=1$ are shown. Trajectories around the turning points $x_{\pm1}$ are also
shown, but not shown is the infinite sequence of U-turns lying between these
trajectories and approaching the negative-real axis!

CMB thanks the U.S.~Department of Energy for financial support. Mathematica 9
was used to produce the numerical calculations and plots in this paper.


\begin{thebibliography}{99}

\bibitem{R1} C.~M.~Bender, S.~Boettcher, and P.~N.~Meisinger,
% "$\cP\cT$-symmetric quantum mechanics"
J.~Math.~Phys.~{\bf 40}, 2201 (1999).

\bibitem{R2} A.~Nanayakkara, Czech.~J.~Phys.~{\bf 54}, 101 (2004).
% "Classical motion of complex 2-D non-Hermitian systems"

\bibitem{R3} A.~Nanayakkara, J.~Phys.~A: Math.~Gen.~{\bf 37}, 4321 (2004).
%Classical trajectories of 1D complex non-Hermitian Hamiltonian systems.

\bibitem{R4} F.~Calogero, D.~Gomez-Ullate, P.~M.~Santini, and M.~Sommacal, 
% "The transition from regular to irregular motions, explained as travel on
% Riemann surfaces"
J.~Phys.~A: Math.~Gen.~{\bf 38}, 8873 (2005).

\bibitem{R5} C.~M.~Bender, J.-H.~Chen, D.~W.~Darg, and K.~A.~Milton, 
% "Classical trajectories for complex Hamiltonians"
J.~Phys.~A: Math.~Gen.~{\bf 39}, 4219 (2006).

\bibitem{R6} Y.~Goldfarb, I.~Degani, and D.~J.~Tannor, J.~Chem.~Phys.~{\bf 125},
231103 (2006); Y.~Goldfarb, and D.~J.~Tannor, {\it Ibid.}~{\bf 127}, 161101
(2007).

\bibitem{R7} C.~M.~Bender and D.~W.~Darg,
% "Spontaneous breaking of classical $\cP\cT$ symmetry"
J.~Math.~Phys.~{\bf 48}, 042703 (2007).

\bibitem{R8} C.~M.~Bender, D.~D.~Holm, and D.~W.~Hook,
% "Complex trajectories of a Simple Pendulum"
J.~Phys.~A: Math.~Theor.~{\bf 40}, F81 (2007)

\bibitem{R9} C.~M.~Bender, D.~D.~Holm, and D.~W.~Hook,
% "Complexified dynamical systems"
J.~Phys.~A: Math.~Theor.~{\bf 40}, F793 (2007).

\bibitem{R10} Y.~Fedorov and D.~Gomez-Ullate,
% "Dynamical systems on infinitely sheeted Riemann surfaces"
Physica D {\bf 227}, 120 (2007).

\bibitem{R11} C.~M.~Bender and D.~W.~Hook, 
% "Exact isospectral pairs of $\cP \cT$-symmetric Hamiltonians"
J.~Phys.~A: Math.~Theor.~{\bf 41}, 244005 (2008).

\bibitem{R12} A.~V.~Smilga, J.~Phys.~A: Math.~Theor.~{\bf 42}, 095301 (2009).
% "Exceptional points in quantum and classical dynamics"

\bibitem{R13} C.~M.~Bender, D.~C.~Brody, and D.~W.~Hook, 
% "Quantum effects in classical systems having complex energy"
J.~Phys.~A: Math.~Theor.~{\bf 41}, 352003 (2008).

\bibitem{R14} Y.~Goldfarb, J.~Schiff, and D.~J.~Tannor, J.~Chem.~Phys.~{\bf
128}, 164114 (2008).

\bibitem{R15} C.~M.~Bender and T.~Arpornthip,
% "Conduction bands in classical periodic potentials"
Pramana J.~Phys.~{\bf 73}, 259 (2009).

\bibitem{R16} C.~M. Bender, J.~Feinberg, D.~W.~Hook, and D.~J.~Weir, Pramana
J.~Phys.~{\bf 73}, 453 (2009).
% "Chaotic Systems in Complex Phase Space" [arXiv: hep-th/0809.1975] 

\bibitem{R17} C.~M.~Bender, D.~W.~Hook, and K.~S.~Kooner,
% "Classical particle in a complex elliptic potential"
J.~Phys.~A: Math.~Theor.~{\bf 43}, 165201 (2010).

\bibitem{R18} C.~M.~Bender, D.~W.~Hook, P.~N.~Meisinger, and Q.~H.~Wang,
% "Complex Correspondence Principle" [arXiv: hep-th/0912.2069] 
Phys.~Rev.~Lett.{\bf 104}, 061601 (2010) and
% "Probability density in the complex plane" [arXiv: hep-th/0912.4659]
Ann.~Phys.~{\bf 325}, 2332 (2010).

\bibitem{R19} A.~G. Anderson, C. M. Bender, and U. I. Morone,
% "Periodic orbits for classical particles having complex energy" 
Phys.~Lett.~A {\bf 375}, 3399 (2011).

\bibitem{R20} A.~Cavaglia, A.~Fring, and B.~Bagchi,
J.~Phys.~A: Math.~Theor.~{\bf 44}, 325201 (2011).
% PT-symmetry breaking in complex nonlinear wave equations & their deformations"

\bibitem{R21} D.~C.~Brody and E.-M.~Graefe,
% "On complexified mechanics and coquaternions"
J.~Phys.~A: Math.~Theor.~{\bf 44}, (2011) 072001.

\bibitem{R22} C.~M.~Bender and D.~W.~Hook,
% "Quantum tunneling as a classical anomaly" 
J.~Phys.~A: Math.~Theor.~{\bf 44}, 372001 (2011).

\bibitem{R23} A.~G.~Anderson and C.~M.~Bender,
% "Complex Trajectories in a Classical Periodic Potential" 
J.~Phys.~A: Math.~Theor.~{\bf 45}, 455101 (2012).

\bibitem{R24} C.~M.~Bender, D.~W.~Hook, and S.~P.~Klevansky,
% "Negative Energy PT-Symmetric Hamiltonians" 
J.~Phys.~A: Math.~Theor.~{\bf 45}, 444003 (2012).

\bibitem{R25} C.~M.~Bender and D.~W.~Hook,
% "Universal Spectral Behavior of $x^2(ix)^\epsilon$ Potentials" 
Phys.~Rev.~A {\bf 86}, 022113 (2012).

\bibitem{R26} N.~Turok, arXiv: quant-ph/1312.1771.

\bibitem{R27} C.~M.~Bender and S.~Boettcher, Phys.~Rev.~Lett.~{\bf 80}, 5243
(1998).
% "Real Spectra in Non-Hermitian Hamiltonians Having {\cal PT} Symmetry"

\bibitem{R28} P.~E.~Dorey, C.~Dunning, and R.~Tateo, J.~Phys.~A: Math.~Gen.~{\bf
34}, L391 (2001) and {\bf 34}, 5679 (2001).

\bibitem{R29} P.~E.~Dorey, C.~Dunning, and R.~Tateo, J.~Phys.~A: Math.~Gen.~{\bf
40}, R205 (2007).
% "The ODE/IM Correspondence" review paper

\bibitem{R30} H.~F.~Jones and J.~Mateo, Phys.~Rev.~D {\bf 73}, 085002 (2006). 

\bibitem{R31} C.~M.~Bender, D.~C.~Brody, J.-H.~Chen, H.~F.~Jones, K.~A.~Milton,
and M.~C.~Ogilvie, Phys.~Rev.~D {\bf 74}, 025016 (2006).

\end{thebibliography}
\end{document}